\documentclass[]{eptcs}
\usepackage{amsmath, amsthm, amssymb, verbatim, mathtools, fancyvrb, varwidth}
\usepackage[all]{xy}
\usepackage{graphicx}
\usepackage{caption}
\usepackage{subcaption}
\usepackage{hhline}
\usepackage{fancybox}

\newenvironment{wideverbatim}%
{\vskip\baselineskip\VerbatimEnvironment
\begin{Sbox}\small\begin{BVerbatim}}
{\end{BVerbatim}%
\end{Sbox}\centerline{\TheSbox}\vskip\baselineskip}

\DeclareFontFamily{OT1}{slmss}{}
\DeclareFontShape{OT1}{slmss}{m}{n}
     {<-8.5> s*[1.1] rm-lmss8
      <8.5-9.5> s*[1.1] rm-lmss9
      <9.5-11> s*[1.1] rm-lmss10
      <11-15.5> s*[1.1] rm-lmss12
      <15.5-> s*[1.1] rm-lmss17
     }{}

\DeclareSymbolFont{sfoperators}{OT1}{slmss}{m}{n}
\DeclareSymbolFontAlphabet{\mathsf}{sfoperators}
\makeatletter
\def\operator@font{\mathgroup\symsfoperators}
\makeatother

\title{Polymorphic Types in ACL2}
\author{Benjamin Selfridge
\institute{University of Texas at Austin\\ Austin, TX}
\email{benself@cs.utexas.edu}
\and
Eric Smith
\institute{Kestrel Institute\\ Palo Alto, CA}
\email{eric.smith@kestrel.edu}
}
\def\titlerunning{Polymorphic Types in ACL2}
\def\authorrunning{B. Selfridge \& E. Smith}
\begin{document}
\maketitle

\begin{abstract}
This paper describes a tool suite for the ACL2 programming language which incorporates certain ideas from the Hindley-Milner paradigm of functional programming (as exemplified in popular languages like ML and Haskell), including a ``typed'' style of programming with the ability to define polymorphic types. These ideas are introduced via macros into the language of ACL2, taking advantage of ACL2's guard-checking mechanism to perform type checking on both function definitions and theorems. Finally, we discuss how these macros were used to implement features of Specware \cite{specware}, a software specification and implementation system.
\end{abstract}

\section{Introduction}
Specware is ``a next-generation environment supporting the design, development and automated synthesis of scalable, correct-by-construction software.'' \cite{specware} The language of Specware is a high-level programming and specification language used to create program specifications, and to refine them into executable code.\footnote{Throughout this paper, we use the term Specware to refer to both the full Specware system, and to Metaslang, the programming language used by Specware.} One of the selling points of a high-powered system like Specware is its robust type system, which accommodates sum type definitions, pattern-matching, and polymorphic types. 

\begin{figure}[t]
\begin{wideverbatim}
type SeqInt =
  | SeqNil
  | SeqCons Int * SeqInt
    
type Seq a = 
  | SeqNil 
  | SeqCons a * (Seq a)
    
op [a] SeqAppend (x:Seq a, y:Seq a) : Seq a =
case x of
  | SeqNil -> y
  | SeqCons (hd,tl) -> SeqCons (hd, SeqAppend (tl, y)) 
    
op [b] SeqRev (x:Seq b) : Seq b =
case x of
  | SeqNil -> SeqNil
  | SeqCons (hd,tl) -> SeqAppend (SeqRev tl, SeqCons (hd,SeqNil))

theorem SeqAppend_Associative is [a]
  fa(x:Seq a,y:Seq a,z:Seq a) 
    SeqAppend(SeqAppend(x,y),z) = SeqAppend(x,SeqAppend(y,z))

theorem SeqAppend_of_SeqNil_1 is [a]
  fa (x:Seq a) SeqAppend(SeqNil,x) = x
theorem SeqAppend_of_SeqNil_2 is [a]
  fa (x:Seq a) SeqAppend(x,SeqNil) = x

theorem SeqRev_of_SeqAppend is [a]
  fa (x:Seq a,y:Seq a) SeqRev (SeqAppend (x,y)) = SeqAppend (SeqRev y, SeqRev x)

theorem SeqRev_of_SeqRev is [a]
  fa (x:Seq a) (SeqRev (SeqRev x)) = x
\end{wideverbatim}
\caption{Our example Specware program.}
\label{ex}
\end{figure}

Figure \ref{ex} contains a snippet of Specware code that will be used as a running example throughout this work. It includes two type definitions, two function definitions (these are called \verb|op|s in Specware), and several theorems. The first type definition listed, \verb|SeqInt|, represents a list of integers. It consists of two type cases: \verb|SeqNil|, representing an empty list, and \verb|SeqCons|, consisting of a integer/\verb|SeqInt| pair. In Specware we call these sum types ``coproduct'' types, and we will stick to this terminology in the remainder of this work.

The next type definition, \verb|Seq a|, is a generalization of the \verb|SeqInt| type to be polymorphic in the type of its contents. We can instantiate the type variable \verb|a| to be any of Specware's built-in base types, or any other type we wish to define. We can also define functions on this new type, like \verb|SeqAppend| and \verb|SeqRev|, and we can state theorems about such functions, like \verb|SeqAppend_Associative| and \verb|SeqRev_of_SeqRev|.

Polymorphic typing greatly enhances the expressibility of Specware, and it is one of many sophisticated language features that ACL2, a first-order untyped programming language and theorem prover, does not support natively. The goal of this work is to present a partial implementation of some of these features in ACL2, which includes a mechanism for defining ``typed'' ACL2 functions and theorems in a style that mimics Specware syntax. This work is some of the fruits of a larger effort to use ACL2 as a back-end prover for Specware. Since Specware emits proof obligations but does not directly verify them, all the proofs must be translated into an external proof environment and verified independently. The ACL2 constructs presented in this work were essential in the translation process, and they also have the potential to be quite useful on their own, independent of Specware. Although our work does not address type inference, this could be achieved by maintaining an ACL2 table mapping functions to types; we discuss this more in the conclusion.

The organization of the remainder of this paper is as follows. Sections 2, 3, and 4 describe how the program in Figure 1 is translated into ACL2 (via the use of some new macros), addressing the coproduct type definitions, function definitions, and theorems respectively. Section 5 concludes the paper by summarizing what has been accomplished and what remains to be done (type inference).

\section{ACL2 ``types'' and polymorphic type definitions}
Before we can begin thinking about defining polymorphic types in ACL2, we must first understand how to implement types in the first place. ACL2 is an untyped language; in the ACL2 logic, every function must accept all types of arguments. However, we can use predicates (functions of arity 1 that return either \verb|T| or \verb|NIL|) to define a type. An example of such a function in ACL2 is \verb|integerp|, which returns \verb|T| if its argument is an integer, and \verb|NIL| otherwise. In this work, we use \verb|Int| to designate this type, and \verb|Int-p| to designate the ACL2 predicate that recognizes \verb|Int|s.\footnote{We define this function as a synonym for \texttt{integerp}, and likewise, we define \texttt{Bool-p} as a synonym for \texttt{booleanp}.} Throughout this work, we maintain a distinction between the name of a type, \verb|Type|, and its recognizer function, \verb|Type-p|. This enables better automation of macros that operate on these types; if the user refers to a type \verb|Type|, the system will automatically append \verb|-p| to its name.

In order to implement coproduct types in ACL2, we started with the pre-existing ACL2 book \verb|defsum|. The \verb|defsum| macro\footnote{This macro was introduced and first used in Swords and Cook \cite{swords}. We added some slight modifications to the original macro in order to accommodate a bookkeeping mechanism that would enable automatic functional instantiation for polymorphic theorems.} uses a combination of guard checking and theorems about the output type of functions (along with many other useful theorems) in order to implement an ad-hoc sum type. It also includes a pattern-matching mechanism, \verb|pm|, which is a more sophisticated version of ACL2's \verb|case| macro. We introduced a new macro, \verb|defcoproduct|, which is defined in terms of \verb|defsum| but also accommodates polymorphism, and a slightly modified version of \verb|pm|, which we named \verb|case-of| in order to more closely reflect the syntax of Specware. We can define the ``concrete'' (non-polymorphic) type \verb|SeqInt| as follows:
\begin{verbatim}
  (defcoproduct SeqInt
    (SeqNil)
    (SeqCons Int SeqInt))
\end{verbatim}
This macro-expands to a simple call to \verb|defsum|:
\begin{verbatim}
  (DEFSUM SEQINT
    (SEQNIL)
    (SEQCONS (INT-P ARG-1) (SEQINT-P ARG-2)))
\end{verbatim}
As we can see, this type is simple enough to have been defined using \verb|defsum| alone. To define a polymorphic \verb|Seq| data structure, however, we need to use \verb|defcoproduct|:
\begin{verbatim}
  (defcoproduct Seq
    :type-vars (a)              ;; type variables
    (SeqNil)                    ;; two type cases - SeqNil and SeqCons
    (SeqCons a (:inst Seq a)))
\end{verbatim}
This code, although its syntax is obviously ACL2, still resembles the original Specware definition of \verb|Seq|. The \verb|defcoproduct| macro defines a new type, \verb|Seq|, by introducing two type constructors, \verb|SeqCons| and \verb|SeqNil|. The type is defined using a single type variable \verb|a|, which is a placeholder for the type of \verb|Seq|'s contents. The \verb|:inst| tag is necessary to inform the macro that we are using a particular instance of the \verb|Seq| type.

Because the logic of ACL2 does not have true polymorphic types (indeed, it does not have a type system at all), the ACL2 definition of \verb|Seq| is not a true type definition, like it is in Specware. Instead, it serves as a template for creating instantiations of \verb|Seq| with a specific type replacing the type variable \verb|a|. The above definition actually introduces a macro, \verb|Seq-instantiate|, which we can use to instantiate \verb|Seq| on integers as follows:
\begin{verbatim}
  (Seq-instantiate int)
\end{verbatim}
This macro-expands (as before) to a call to \verb|defsum|:
\begin{verbatim}
  (DEFSUM SEQ-INT
    (SEQNIL-INT)
    (SEQCONS-INT (INT-P ARG-1)
                 (SEQ-INT-P ARG-2)))
\end{verbatim}
We can see this looks nearly identical to the definition of the concrete coproduct type \verb|SeqInt| above. (We have deliberately omitted some bookkeeping information from this \verb|defsum| call - this information will be necessary when we wish to instantiate theorems about the \verb|Seq| type on specific instances, but the details of this are not important.)

We can also define polymorphic types with more than one type variable, or in terms of previously defined polymorphic types:
\begin{verbatim}
  (defcoproduct EitherSeq
    :type-vars (a b)
    (LeftSeq  (:inst Seq a))
    (RightSeq (:inst Seq b)))
\end{verbatim}
This defines a new polymorphic type, \verb|EitherSeq|, parameterized by variables \verb|a| and \verb|b|. We can now instantiate it with concrete types:
\begin{verbatim}
(EitherSeq-instantiate int bool)
\end{verbatim}
The above call expands to
\begin{verbatim}
  (PROGN (SEQ-INSTANTIATE INT)
         (SEQ-INSTANTIATE BOOL)
         (DEFSUM EITHERSEQ-INT-BOOL
           (LEFTSEQ-INT-BOOL (SEQ-INT-P ARG-1))
           (RIGHTSEQ-INT-BOOL (SEQ-BOOL-P ARG-1))))
\end{verbatim}
Notice that before \verb|EitherSeq| is instantiated on integers and booleans, the \verb|Seq| type must be instantiated on these two types. This is automatically checked by the \verb|defcoproduct| macro, and these instantiations are collected and included before defining \verb|EitherSeq-Int-Bool|. It is notable that none of the macros presented in this paper query the ACL2 world. If some types have already been defined, the preliminary type instantiations will be dismissed by ACL2 as redundant definitions, and the macro will still succeed.

The polymorphism supported by \verb|defcoproduct| is somewhat limited; the macro does not support mutually recursive datatypes, and all instantiation must happen in one step (i.e., we cannot instantiate the \verb|a| variable of \verb|EitherSeq| and leave \verb|b| as a type variable). However, these are not fundamental limitations, and could be implemented with more work.

\section{Polymorphic functions and theorems}
Consider the Specware function \verb|SeqAppend| defined in Figure \ref{ex}. This is the ``append'' function on lists, defined for the polymorphic type \verb|Seq|. We can translate this definition into ACL2 using our new \verb|defun-typed| macro:
\begin{verbatim}
  (defun-typed SeqAppend
    :type-vars (a)                         ;; type variables
    ((x (:inst Seq a)) (y (:inst Seq a)))  ;; typed argument list
    (:inst Seq a)                          ;; output type
    (case-of x                             ;; function body
      (((:inst SeqNil a))                  ;; case 1: SeqNil
       y)
      (((:inst SeqCons a) hd tl)           ;; case 2: SeqCons
       ((:inst SeqCons a) hd ((:inst SeqAppend a) tl y)))))
\end{verbatim}
The \verb|defun-typed| macro is a version of \verb|defun| that requires type annotations for all its input values, as well as a type annotation for its own return value. We supply a list of type variables (which can be omitted if there are none), a list of the arguments with their associated types, an output type, and the body of the function. 

One obvious weakness of this definition is its verbosity. Every time we pattern match on a polymorphic type, we must include the \verb|:inst| keyword in each pattern to indicate which instantiation for a given constructor we are using. This could, of course, be solved by implementing type inference; for our immediate task of translating Specware code to ACL2, this was not necessary, but it could be done with some more work.

In order to use this function, we must first instantiate it:
\begin{verbatim}
(SeqAppend-instantiate int)
\end{verbatim}
This macro-expands to
\begin{verbatim}
 (PROGN (SEQ-INSTANTIATE INT)
        (SEQNIL-INSTANTIATE INT)
        (SEQCONS-INSTANTIATE INT)
        (DEFUN SEQAPPEND-INT (X Y)
           (DECLARE (XARGS :GUARD (AND (SEQ-INT-P X) (SEQ-INT-P Y))
                           :VERIFY-GUARDS NIL))
           (IF (MBT (AND (SEQ-INT-P X) (SEQ-INT-P Y)))
               (CASE-OF X ((SEQNIL-INT) Y)
                        ((SEQCONS-INT HD TL)
                         (SEQCONS-INT HD (SEQAPPEND-INT TL Y))))
               NIL))
        (DEFTHM SEQAPPEND-INT-TYPE
                (IMPLIES (AND (SEQ-INT-P X) (SEQ-INT-P Y))
                         (SEQ-INT-P (SEQAPPEND-INT X Y)))
                :RULE-CLASSES (:TYPE-PRESCRIPTION :REWRITE))
        (VERIFY-GUARDS SEQAPPEND-INT)
\end{verbatim}
As before, we first instantiate the polymorphic \verb|Seq| type for \verb|int|s before defining our \verb|SeqAppend-Int| function. The \verb|SEQ-INSTANTIATE|, \verb|SEQNIL-INSTANTIATE|, and \verb|SEQCONS-INSTANTIATE| are all redundant; they have the exact same definition. The \verb|defun-typed| macro scans its argument list, output type, and body for the \verb|:inst| keyword, and calls the associated instantiation macro for each occurrence. This was a brute-force way to guarantee that all the necessary functions and types will be defined before the current function definition is submitted. Notice how a combination of ACL2 guards and theorems are used to ensure that \verb|SeqAppend-Int| satisfies all the typing requirements; we require that the guards of the function calls made in the definition of \verb|SeqAppend-Int| are never violated given our assumptions about the types of \verb|x| and \verb|y|, and we also require that, assuming both input variables are \verb|Seq-Int|s, the output of this function is also a \verb|SeqInt|. Notice how we check the latter first; for recursive definitions, it is often the case that we need to know the output type of the function before we can verify the guards.

Of course, we can also define polymorphic functions in terms of other, previously defined ones.
\begin{verbatim}
(defun-typed SeqRev
  :type-vars (a)
  ((x (:inst Seq a)))
  (:inst Seq a)
  (case-of x
    (((:inst SeqNil a)) ((:inst SeqNil a)))
    (((:inst SeqCons a) hd tl) 
     ((:inst SeqAppend a) 
      ((:inst SeqRev a) tl) 
      ((:inst SeqCons a) hd ((:inst SeqNil a)))))))
\end{verbatim}
If we instantiate \verb|SeqRev| with the concrete type \verb|bool| via
\begin{verbatim}
(SeqRev-instantiate bool)
\end{verbatim}
this will macro-expand to
\begin{verbatim}
(PROGN
  (SEQ-INSTANTIATE BOOL)
  (SEQAPPEND-INSTANTIATE BOOL)
  (SEQCONS-INSTANTIATE BOOL)
  (SEQNIL-INSTANTIATE BOOL)
  (DEFUN SEQREV-BOOL (X)
    (DECLARE (XARGS :GUARD (SEQ-BOOL-P X)
                    :VERIFY-GUARDS NIL))
    (IF (MBT (SEQ-BOOL-P X))
        (CASE-OF X ((SEQNIL-BOOL) (SEQNIL-BOOL))
                   ((SEQCONS-BOOL HD TL)
                    (SEQAPPEND-BOOL (SEQREV-BOOL TL)
                                    (SEQCONS-BOOL HD (SEQNIL-BOOL)))))
        NIL))
  (DEFTHM SEQREV-BOOL-TYPE
          (IMPLIES (SEQ-BOOL-P X)
                   (SEQ-BOOL-P (SEQREV-BOOL X)))
          :RULE-CLASSES (:TYPE-PRESCRIPTION :REWRITE))
  (VERIFY-GUARDS SEQREV-BOOL))
\end{verbatim}
Notice that both the \verb|Seq| type and the \verb|SeqAppend| function are instantiated for \verb|bool| before defining \verb|SeqRev-Bool|. Of course, it would have sufficed to only invoke \verb|(SEQAPPEND-INSTANTIATE BOOL)|, but our \verb|defun-typed| macro is not smart enough to figure that out; everywhere it sees an \verb|:inst| keyword, it calls the associated instantiation macro.

We can also state (and prove) theorems about functions involving polymorphic types using our new macro \verb|defthm-typed|. For instance, we can translate the \verb|SeqAppend_Associative| theorem from the introduction into ACL2 like so:
\begin{verbatim}
  (defthm-typed SeqAppend_Associative
    :type-vars (a)                     ;; type variables
    ((x (:inst Seq a))                 ;; type annotations for free variables
     (y (:inst Seq a))
     (z (:inst Seq a)))
    (equal                             ;; theorem body
     ((:inst SeqAppend a) ((:inst SeqAppend a) x y) z)   
     ((:inst SeqAppend a) x ((:inst SeqAppend a) y z))))
\end{verbatim}
This macro-expands to
\begin{verbatim}
 (PROGN
  (ENCAPSULATE (((A-P *) => *))
               (LOCAL (DEFUN A-P (X) (DECLARE (IGNORE X)) T))
               (DEFTHM A-TYPE (BOOLEANP (A-P X))
                       :RULE-CLASSES :TYPE-PRESCRIPTION))
  (SEQ-INSTANTIATE A)
  (SEQAPPEND-INSTANTIATE A)
  (DEFUND-TYPED SEQAPPEND_ASSOCIATIVE-A-BODY
    ((X SEQ-A) (Y SEQ-A) (Z SEQ-A))
    BOOL
    (EQUAL (SEQAPPEND-A (SEQAPPEND-A X Y) Z)
           (SEQAPPEND-A X (SEQAPPEND-A Y Z))))
  (DEFTHM SEQAPPEND_ASSOCIATIVE-A
          (IMPLIES (AND (SEQ-A-P X)
                        (SEQ-A-P Y)
                        (SEQ-A-P Z))
                   (EQUAL (SEQAPPEND-A (SEQAPPEND-A X Y) Z)
                          (SEQAPPEND-A X (SEQAPPEND-A Y Z)))))
  (DEFMACRO SEQAPPEND_ASSOCIATIVE-INSTANTIATE (A)
   ;; ... macro definition omitted
    )
\end{verbatim}
The \verb|defthm-typed| macro does several things. First, it defines an encapsulated predicate, \verb|A-P|, which will be used to represent the type variable \verb|a|. Then, after instantiating all the needed types and functions, we type check the body of the theorem by defining it as a function with output type \verb|bool| (if the theorem doesn't even type check, then we don't need to bother to try and prove it). Then, it proves a version of \verb|SeqAppend_Associative| where the \verb|Seq| type has been instantiated on an encapsulated predicate \verb|A-P|. In theory, this proves the theorem in general for any type instantiation. Finally, a new macro, \verb|SEQAPPEND_ASSOCIATIVE-INSTANTIATE|, is introduced, which allows us to prove this theorem for a specific instantiation of the \verb|Seq| type. This macro uses functional instantiation (along with a substantial amount of bookkeeping) to prove the theorem automatically from the original theorem, \verb|SEQAPPEND_ASSOCIATIVE-A|. If we instantiate this theorem for integers via
\begin{verbatim}
  (SeqAppend_Associative-instantiate int)
\end{verbatim}
we get
\begin{verbatim}
  (PROGN
   (SEQ-INSTANTIATE INT)
   (SEQAPPEND-INSTANTIATE INT)
   (DEFTHM-TYPED
       SEQAPPEND_ASSOCIATIVE-INT
       ((X SEQ-INT) (Y SEQ-INT) (Z SEQ-INT))
       (EQUAL (SEQAPPEND-INT (SEQAPPEND-INT X Y) Z)
              (SEQAPPEND-INT X (SEQAPPEND-INT Y Z)))
       :HINTS
       (("Goal" :DO-NOT-INDUCT T
                :IN-THEORY (ENABLE SEQ-INT-FUNCTIONS)
                :USE ((:FUNCTIONAL-INSTANCE 
                       SEQAPPEND_ASSOCIATIVE-A (A-P INT-P)
                       (SEQ-A-P SEQ-INT-P)
                       (SEQCONS-A-P SEQCONS-INT-P)
                       (SEQNIL-A-P SEQNIL-INT-P)
                       (SEQCONS-A SEQCONS-INT)
                       (SEQNIL-A SEQNIL-INT)
                       (SEQCONS-A-ARG-2 SEQCONS-INT-ARG-2)
                       (SEQCONS-A-ARG-1 SEQCONS-INT-ARG-1)
                       (SEQAPPEND-A SEQAPPEND-INT)))))))
\end{verbatim}
Notice how the theorem is proved using functional instantiation on the more general theorem, \\\verb|SeqAppend_Associative-A|.

We can use the macros described above to implement the entire Specware program of Figure \ref{ex} using our four new macros, \verb|defcoproduct|, \verb|defun-typed|, \verb|defthm-typed|, and \verb|case-of|. The full listing for the ACL2 version of this program is given in Figure \ref{ex_acl2}.

\begin{figure}[ht]
\begin{wideverbatim}
(in-package "ACL2")
(include-book "~/Desktop/specware-files/code/specware-book")
(set-ignore-ok t)
(set-bogus-defun-hints-ok t)

(defcoproduct SeqInt
  (SeqNil)
  (SeqCons Int SeqInt))
  
(defcoproduct Seq
  :type-vars (a)
  (SeqCons a (:inst Seq a))
  (SeqNil))

(defun-typed SeqAppend
  :type-vars (a)
  ((x (:inst Seq a)) (y (:inst Seq a)))
  (:inst Seq a)
  (case-of x
    (((:inst SeqNil a)) y)
    (((:inst SeqCons a) hd tl)
     ((:inst SeqCons a) hd ((:inst SeqAppend a) tl y)))))

(defthm-typed SeqAppend_Associative
  :type-vars (a)
  ((x (:inst Seq a))
   (y (:inst Seq a))
   (z (:inst Seq a)))
  (equal ((:inst SeqAppend a) ((:inst SeqAppend a) x y) z) 
         ((:inst SeqAppend a) x ((:inst SeqAppend a) y z))))

(defthm-typed SeqAppend_of_SeqNil_1
  :type-vars (a)
  ((x (:inst Seq a)))
  (equal ((:inst SeqAppend a) ((:inst SeqNil a)) x) x))
(defthm-typed SeqAppend_of_SeqNil_2
  :type-vars (a)
  ((x (:inst Seq a)))
  (equal ((:inst SeqAppend a) x ((:inst SeqNil a))) x))
\end{wideverbatim}
\end{figure}
\begin{figure}[t]
\begin{wideverbatim}
(defun-typed SeqRev
  :type-vars (a)
  ((x (:inst Seq a)))
  (:inst Seq a)
  (case-of x
    (((:inst SeqNil a)) ((:inst SeqNil a)))
    (((:inst SeqCons a) hd tl) 
     ((:inst SeqAppend a) 
      ((:inst SeqRev a) tl) 
      ((:inst SeqCons a) hd ((:inst SeqNil a)))))))
      
(defthm-typed SeqRev_of_SeqAppend
  :type-vars (a)
  ((x (:inst Seq a))
   (y (:inst Seq a)))
  (equal ((:inst SeqRev a) ((:inst SeqAppend a) x y)) 
          ((:inst SeqAppend a) ((:inst SeqRev a) y) 
                               ((:inst SeqRev a) x))))

(defthm-typed SeqRev_of_SeqRev
  :type-vars (a)
  ((x (:inst Seq a)))
  (equal ((:inst SeqRev a) ((:inst SeqRev a) x)) x)
  :hints (("Goal" :in-theory (enable SeqAppend-a SeqRev-a))))
\end{wideverbatim}
\caption{Our example Specware program, defined in ACL2.}
\label{ex_acl2}
\end{figure}

\section{Conclusions}
These macros were introduced in an ACL2 book in order to facilitate the translation process from Specware to ACL2. Instead of having our translator produce the raw ACL2 code, we hide the messy details of implementing these high-level language features with ACL2 macros, which has the advantage of both making the translation process easier and making the automatically generated ACL2 code much more readable. The \verb|gen-acl2| tool was added to Specware in order to automatically translate Specware programs into ACL2 code that uses the macros defined here (the ACL2 code in Figure \ref{ex_acl2} was generated by this tool). 

A byproduct of this effort was the ACL2 macros themselves, which are quite useful in their own right, and suggest the possibility of implementing more of these high-level features in ACL2. Limitations of the polymorphism presented here include the inability to define mutually recursive polymorphic types, as well as the lack of partial type instantiation. These macros could be extended in a straightforward way to include these  features. 

Type inference could be implemented by maintaining an ACL2 table mapping function names to their types (essentially, the ``typing'' theorems exported by the \verb|defun-typed| macro). When the user defines a new function, the Hindley-Milner algorithm can be used to deduce the necessary input and output types of the function (assuming all the function calls in the body already exist in the table), and we can export a theorem capturing this which could be proved in a theory that only includes the typing theorems of previously defined functions. 

We also believe that the techniques used in this paper could be used to introduce ``higher-order'' functions; in particular, we can modify our technique of extracting away a type variable by introducing an encapsulated predicate, to extracting away a function by introducing an encapsulated function. We have not thoroughly investigated the viability of this idea, but it seems to us like a fruitful avenue for further research.

\clearpage
\bibliographystyle{eptcs}
\bibliography{acl2poly}

\end{document}